# LIKE-SIGN CHARGED HIGGS BOSON PRODUCTION IN $e^-e^-$ COLLISIONS AT THE NLC [*]

THOMAS G. RIZZO [†]

*Stanford Linear Accelerator Center, Stanford University, Stanford, CA 94309, USA*

## ABSTRACT

We consider the production of a pair of like-sign charged Higgs bosons in $e^-e^-$ collisions at the NLC within the context of several electroweak models with extended symmetry breaking sectors. We find that the rate for this process, which proceeds through $W^-W^-$ fusion, is a very sensitive probe of the nature of these extended Higgs sectors and that the corresponding cross sections can vary by as much as several orders of magnitude at NLC energies.

The scalar sector of the Standard Model(SM) which is responsible for the spontaneous breaking of $SU(2)_L \otimes U(1)_Y$ consists of only a single $SU(2)_L$ doublet. This results in there being only one real physical Higgs field after the Goldstone bosons are eaten by the $W$ and $Z$. As is well-known, this rather simple situation is far from unique[1] even when we impose the naturalness condition that $\rho = M_W^2/M_Z^2 c_w^2 = 1$ at the tree level, as we must to maintain agreement with high precision electroweak data[2]. Of course in models where the Higgs sector consists solely of $SU(2)_L$ doublets and singlets the condition that $\rho_{tree} = 1$ is naturally satisfied automatically. If the Higgs sector is extended to triplet or other higher dimensional representations, this relation is generally no longer natural without the imposition of additional symmetries. We note as a caveat, however, that it is possible that other rather bizarre Higgs representations may be present in addition to doublets and singlets, and also yield $\rho_{tree} = 1$ *without* the imposition of additional symmetries. These scenarios are discussed separately by Gunion in these Proceedings[3]. A common feature of almost all extended Higgs sectors is the existence of at least one singly-charged Higgs scalar, $H^\pm$. The only semi-model independent limit on the mass of the charged Higgs, $\simeq 45$ GeV, arises from LEP and SLD data on the properties and decay modes[2] of the $Z$, but even this bound must be used carefully.

In $e^+e^-$ collisions, $H^\pm$ is pair produced in the usual fashion, via $s$-channel $\gamma, Z$ exchange, provided this is kinematically allowed. Unfortunately, this production process tells us little about the $H^\pm$ since the corresponding cross section depends only on the electric charge and third component of weak isospin of $H^\pm$, the first of which is fixed. Of course, by studying $H^\pm$ decay modes we may learn something about its relative couplings to fermions and gauge bosons. On the otherhand, $e^-e^-$ collisions at NLC energies allow us to probe these gauge interactions directly through the production process in a rather unique manner. Here in $e^-e^-$, $H^\pm$ is produced via $WW$-fusion subprocess, $W^-W^- \to H^-H^-$, where the $W$'s are radiated off of the initial $e^-$ lines. Thus, like sign Higgs production actually corresponds to the reaction $e^-e^- \to H^-H^-\nu\nu$.

[*]Work supported by the Department of Energy, contract DE-AC03-76SF00515.
[†]Presented at the *Santa Cruz Workshop on $e^-e^-$ Physics at the NLC*, University of California, Santa Cruz, September 4-5, 1995.

The simplest model which allows for $W^-W^- \to H^-H^-$ subprocess is clearly the Two Higgs Doublet Model(THDM), where the relevant gauge and Higgs interactions are completely determined by gauge invariance and are independent of how the scalar sector couples to fermions[5]. In the THDM, the process of interest takes place through $t$- and $u$-channel neutral Higgs exchange, $i.e.$, through the $W(h, H, A)H^\pm$ couplings, where as usual $h(H)$ denotes the light(heavy) CP-even scalar and $A$ is CP-odd. (We will neglect the possibility of CP violating mixings in the Higgs sector for simplicity in the discussion below. It will have no influence on the generality of our conclusions.) In this case it is easy to show that the $t$-channel amplitude for the $W^-W^- \to H^-H^-$ process is proportional to the quantity

$$A_t = \frac{cos^2(\alpha - \beta)}{t - m_h^2} + \frac{sin^2(\alpha - \beta)}{t - m_H^2} - \frac{1}{t - m_A^2}, \qquad (1)$$

with a similar result for the $u$-channel amplitude, $A_u$, which is obtained from $A_t$ via the replacement $t \to u$ in the above expression. As usual, the angle $\alpha$ is responsible for the diagonalization of the CP-even mass matrix, whereas $tan\beta = v_1/v_2$ is the ratio of the two Higgs doublet vacuum expectation values. The expression for $A_t$ has a number of interesting features. First, we observe that $A$ exchange destructively interferes with $h, H$ exchange due to its opposite CP properties. Second, we observe that in the limit that all three neutral Higgs are degenerate the amplitude vanishes. This means that in certain 'symmetry limits' we should expect the $W^-W^- \to H^-H^-$ subprocess cross section to be very small.

In the THDM context, the most 'symmetric' situation is realized in its tree-level supersymmetric version. In that case, specifying $tan\beta$ and $m_{H^+}$ completely determines all of the masses of the the neutral Higgs fields as well as the mixing angle $\alpha$. Let us consider two extreme cases for the values of $tan\beta$ and see what happens. For $tan\beta = 1$, $h$ is massless, $cos(\alpha - \beta) = 0$, $m_H^2 = m_{H^+}^2 + M_Z^2 - M_W^2$, and $m_A^2 = m_{H^+}^2 + M_W^2$. This implies $A_t$ is small and that a very strong cancellation in $A_t$ occurs as $m_{H^+}$ increases. In the opposite limit, $tan\beta \to \infty$, we find instead that $cos(\alpha - \beta) \to 0$ while $M_H \to M_A$ which again leads to a rapid cancellation in the amplitudes. We thus expect on rather general grounds that at the tree level in the SUSY THDM, the like-sign Higgs production cross section is very small and this is explicitly displayed in Figs. 1 and 2. In fact, these cross sections are just too small to be observed at the NLC even with a healthy integrated luminosity. Of course, tree level SUSY is an incomplete theory due to the existence of significant radiative corrections on account of the large logs involving the top quark mass as well as the SUSY partner masses. These corrections destroy the rather simple tree level relationships between the mixing angles and Higgs masses which then results in a significant softening of the cancellations taking place in $A_{t,u}$. Thus we anticipate a large increase in the cross section for $H^-H^-$ in this case depending upon the effectiveness of the radiative corrections. Figs. 1 and 2 show that these radiative contributions lead to a drastic increase in the cross section, even to values of order 0.1 fb, which may be sufficiently large as to lead to an observable signal for very high luminosities. For purposes of demonstration we have chosen the values $\mu = -250$ GeV, $m_0 = 1$ TeV, and $m_t = 180$ GeV with equal $A$ parameters of 500 GeV for the $t$ and $b$ quark sectors to perform our corrections. We have checked that the loop corrected cross sections are not very sensitive to this particular choice.

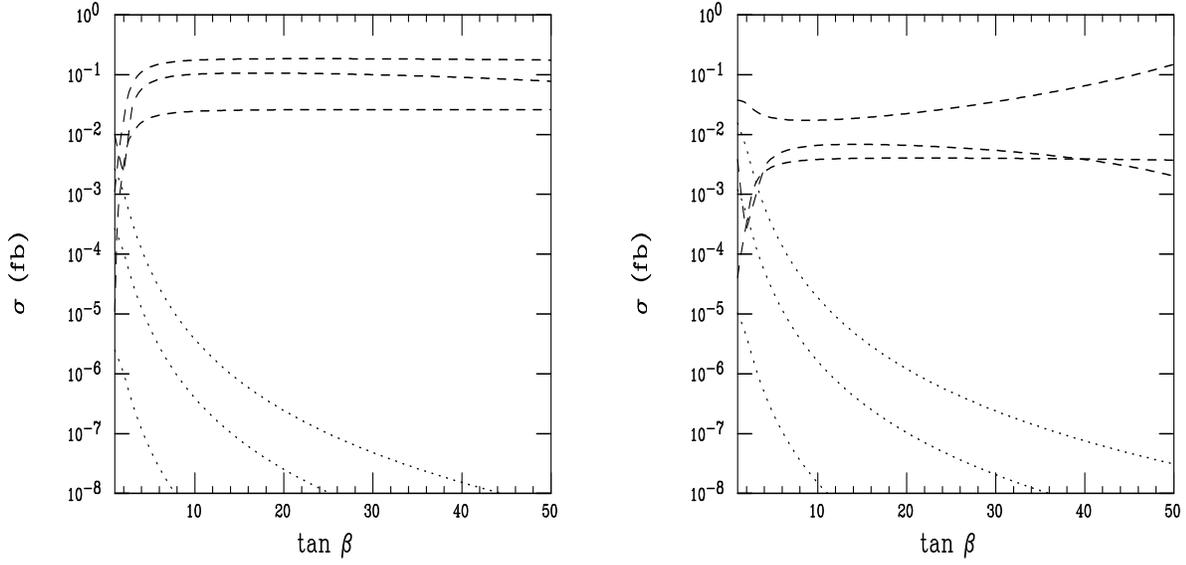

Fig. 1. Like-sign production cross section in the SUSY THDM as a function of $tan\beta$. In (a) for a 2 TeV NLC and in (b) for a 1 TeV NLC for SUSY without(with) radiative corrections corresponding to the dotted(dashed) curves. In (a) the charged Higgs masses for the dotted(dashed) curves are, from top to bottom, 250, 400, and 700(400, 250, and 700) GeV, respectively. In (b) the corresponding masses are 125, 200, and 350 GeV for both sets of curves.

Of course, our results as so far presented are still within the SUSY framework. The region of parameter space occupied by the SUSY THDM is only a very small fraction of the more general case. In a *general* THDM without any of the SUSY restrictions we might expect the cross sections to be allowed to be larger still since there are no *a priori* cancellations. In Fig. 3 we show that this is indeed the case for judiciously chosen values of the THDM parameters with cross sections surpassing 1 fb in some cases. It is clear from the above analysis that the THDM allows for a very wide range values for the of cross section for the process $e^-e^- \to H^-H^-\nu\nu$ but they can never be larger than a few $fb$. To get larger rates we must go beyond the THDM. Adding additional doublets, while increasing the number of parameters substantially, will clearly not lead to enhanced cross sections. By similar arguments it is easy to show that the addition of Higgs singlets will not work either. Thus we are left to consider the possibility of higher dimensional scalar representations.

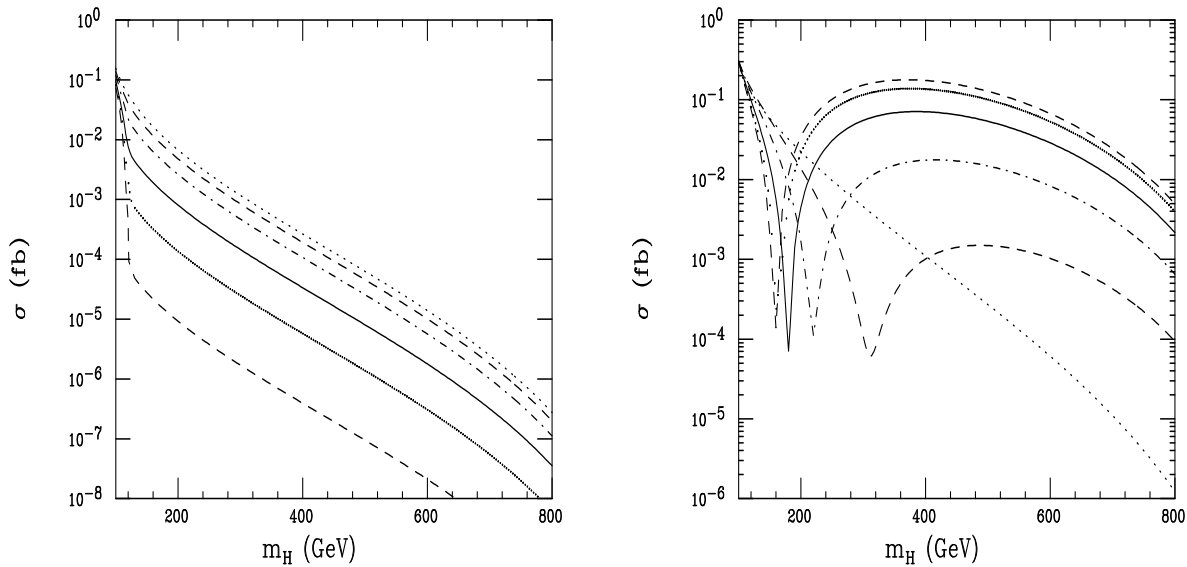

Fig. 2. Like-sign production cross section in the SUSY THDM as a function of the charged Higgs mass for a 2 TeV NLC. In (a) no radiative corrections are included and from top to bottom the curves correspond to $tan\beta$ =1, 1.5, 2, 3, 5 and 10. In (b) radiative corrections are included and run in reverse order in $\tan\beta$ on the right-hand side of the plot.

As discussed above, going beyond Higgs doublets and singlets can be a dangerous proposition if we want to maintain $\rho_{tree} = 1$. The simplest model which satisfies this relation is due to Georgi and Machacek[6] and has been discussed at some length in the recent literature[7]. The Higgs weak eigenstates in this model consists of the usual SM doublet together with a real triplet and a pair of conjugate triplets. An additional global symmetry is imposed on the scalar potential so that the triplet vev's are identical. After spontaneous symmetry breaking, the remaining physical fields can be classified by their custodial $SU(2)$ transformation properties and are degenerate within each self-conjugate multiplet. These fields consist of a **5**, which contains a neutral field as well as a singly and doubly charged pair, a **3** containing a CP-odd neutral field together with a charged pair, as well as two additional neutral singlets **1** and **1'**. Ignoring the possibility of **1-1'** mixing, this leads to four mass parameters: $m_{5,3,1,1'}$. One additional parameter is also present, $tan\phi_H(t_H)$, which is the ratio of the triplet to doublet vev's so that the SM is obtained in the $t_H \to 0$ limit. (We denote $c_H = cos\phi_H$ and $s_H = sin\phi_H$ in what follows.) For simplicity of discussion below, we will assume that $m_1 = m_{1'}$ in order to reduce the size of the parameter space to be explored. The physical spectrum of this model is so rich that it now allows for *both* $W^-W^- \to H_{3,5}^- H_{3,5}^-$ production, but we will see that the characteristics of these two processes are quite distinct.

Let us first consider the $W^-W^- \to H_5^- H_5^-$ process which goes through the $u$- and $t$-channel neutral $H_{3,5}$ exchanges in analogy with what we found in the case of the THDM. However, due to the existence of a $W^+ZH_5^-$ vertex with a strength proportional to $s_H$, there is now also a $Z$ boson exchange diagram. This, however, does not lead to any additional parameter dependence, i.e., the cross section depends only on the variables $m_{3,5}$ and $c_H^2$. Fig.4 shows the $e^-e^- \to H_5^-H_5^-\nu\nu$ cross section for several parameters choices. First, we see that the rate is only mildly dependent on $m_3$ whereas the $c_H^2$ dependence is substantially stronger. In addition, note that $H_3^0$ totally decouples when $c_H^2 = 0$. Second, there is a very strong $m_5$ dependence which enters both through the mass of the final state charged Higgs field as well as through the $H_5^0$ which is exchanged in the $t$- and $u$-channels. Third, and perhaps most interesting, is the fact that the cross sections can now be larger than 10 fb, a value substantially higher than what can be obtained from the THDM.

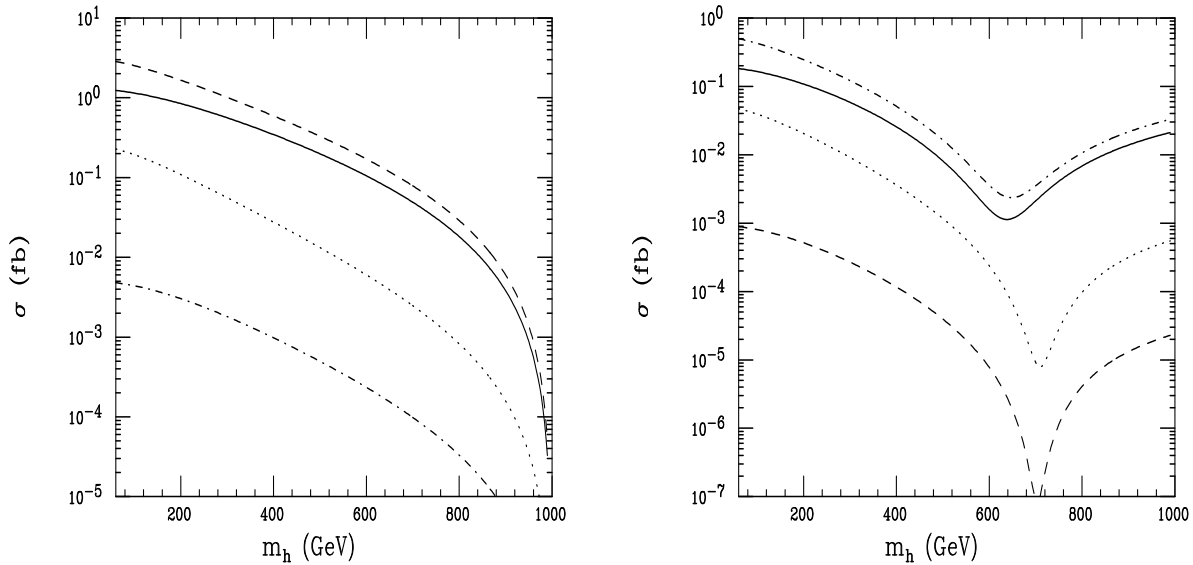

Fig. 3. Like-sign production cross section in the general THDM as a function of the light neutral Higgs mass. In both figures, the upper(lower) pair of curves correspond to charged Higgs masses of 250 and 400 GeV at a 2(1) TeV NLC. In (a) we assume $m_H$ is arbitrary and $cos^2(\alpha - \beta) = 1$ whereas in (b) $m_H$=2 TeV and $cos^2(\alpha - \beta) = 0.5$. $m_A = 1$ TeV in both cases.

The process $e^-e^- \to H_3^-H_3^-\nu\nu$ is even more exotic although no $Z$ exchange is involved; in this case *all* of the four neutral Higgs fields are exchanged in the $t$- and $u$-channels. In addition, due to the existence of a $H_5H_3H_3$ as well as a $W^-W^-H_5^=$ coupling, there is now a contribution due to $s$-channel $H_5^=$ exchange. Interestingly, since it possible that $m_5 > 2m_3$, an $s$-channel resonance may now be present which can lead to a substantial cross section increase. Unfortunately, in this case the cross section now depends upon *all* of the parameters of the model which makes a complete and detailed analysis extremely difficult.

Fig.5 shows an example of non-resonant like-sign $H_3^-$ pair production. We see that the cross section is generally comparable to the case of $H_5^-$ discussed previously-generally in the range of a few to 10's of fb. The cross section is found to be rather sensitive to both the values of $m_1$ and $c_H^2$, falling off quite rapidly with increasing $m_3$. There is potentially a strong destructive interference in this case due to $H_3^0$ exchange since it is CP-odd and turning off $c_H$ decouples the contributions of both $H_{1',5}^0$. The more interesting situation in the case of like-sign $H_3^-$ pair production is when $m_5$ is large. Fig.6 shows that the pair cross section grows dramatically as the resonance contribution is slowly turned on. For small values of $m_5$, we see rates of only a few fb, but once the resonance threshold is passed these can grow dramatically to 100 fb or more. Such a large cross section should be easily observable at the NLC and a like sign mass distribution will clearly show the doubly charged Higgs peak.

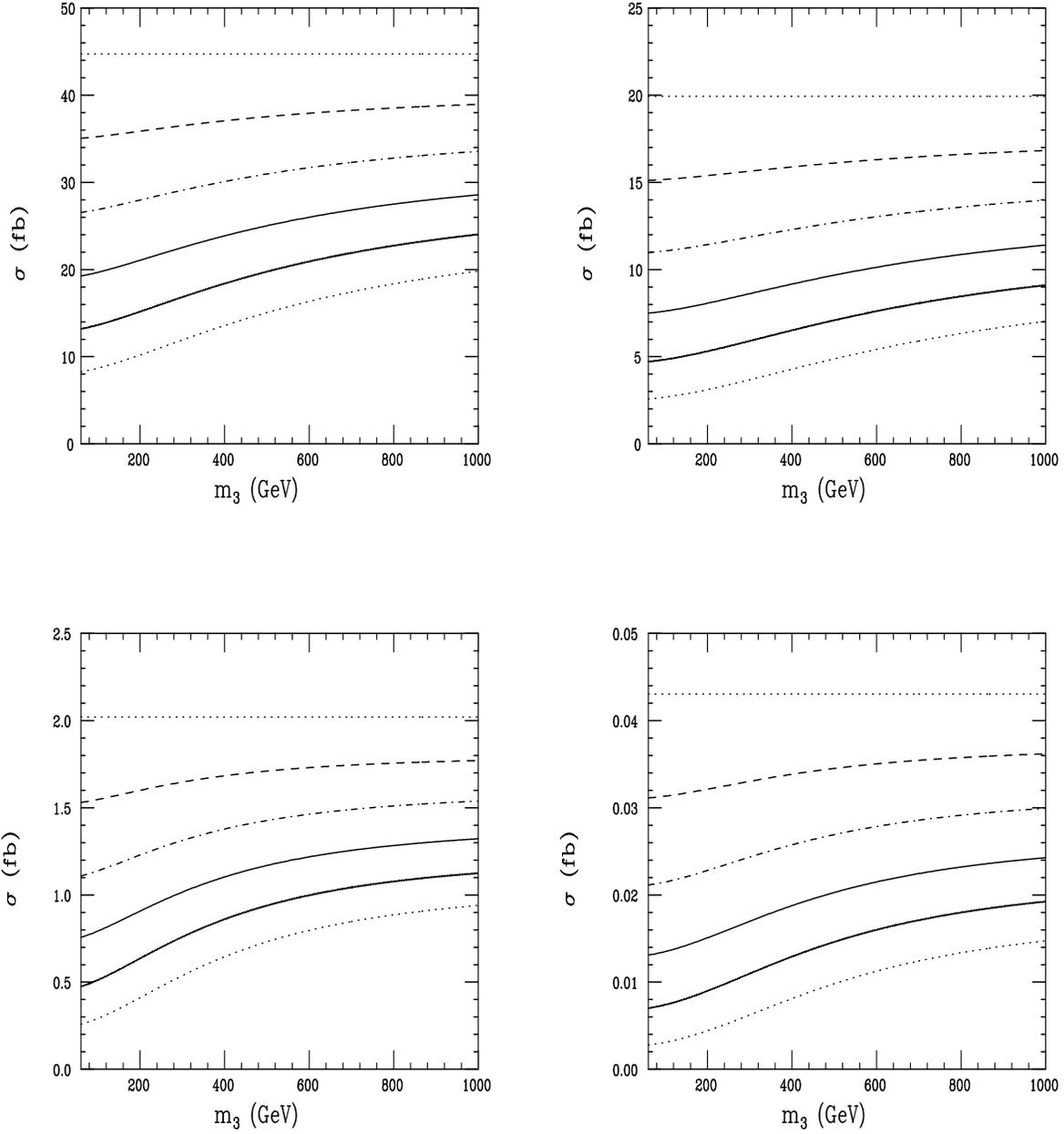

Fig. 4. $e^-e^- \to H_5^- H_5^- \nu\nu$ production cross sections as functions of $m_3$ with $c_H^2$ varying from 0 to 1 in steps of 0.2 from top to bottom in each plot. A 2[1] TeV NLC is assumed in (a) and (b)[(c) and (d)] and $m_5$ is taken to be 400[250] GeV in (a) and (c)[(b) and (d)].

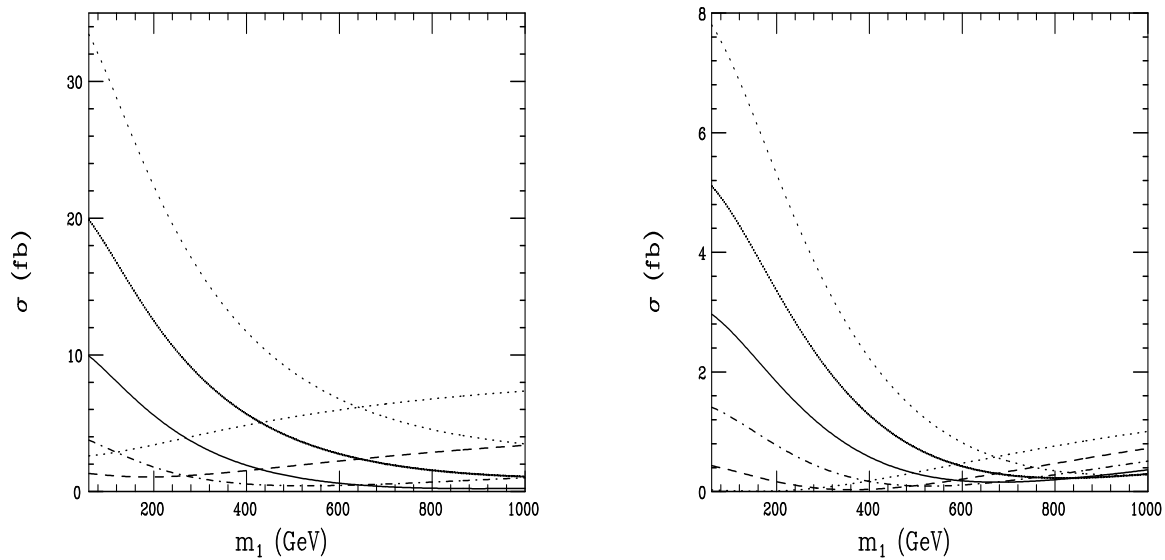

Fig. 5. $e^-e^- \to H_3^- H_3^- \nu\nu$ production cross sections as functions of $m_1 = m_{1'}$ with $c_H^2$ varying from 1 to 0 in steps of 0.2 from top to bottom on the left side of each figure. A 2 TeV NLC and $m_5 = 500$ GeV is assumed in both cases. In (a)[(b)] $m_3 = 250[400]$ GeV is also assumed.

In summary, we have seen how the cross section for the process $e^-e^- \to H^-H^-\nu\nu$ is extremely sensitive to the details of the scalar sector of extended models. Within the THDM alone we found that the cross section was allowed to vary over seven orders of magnitude. In models with triplets with additional global symmetries, the cross sections were allowed to be larger still by two orders of magnitude. Clearly, the $e^-e^- \to H^-H^-\nu\nu$ process is an excellent probe of the symmetry breaking sector.

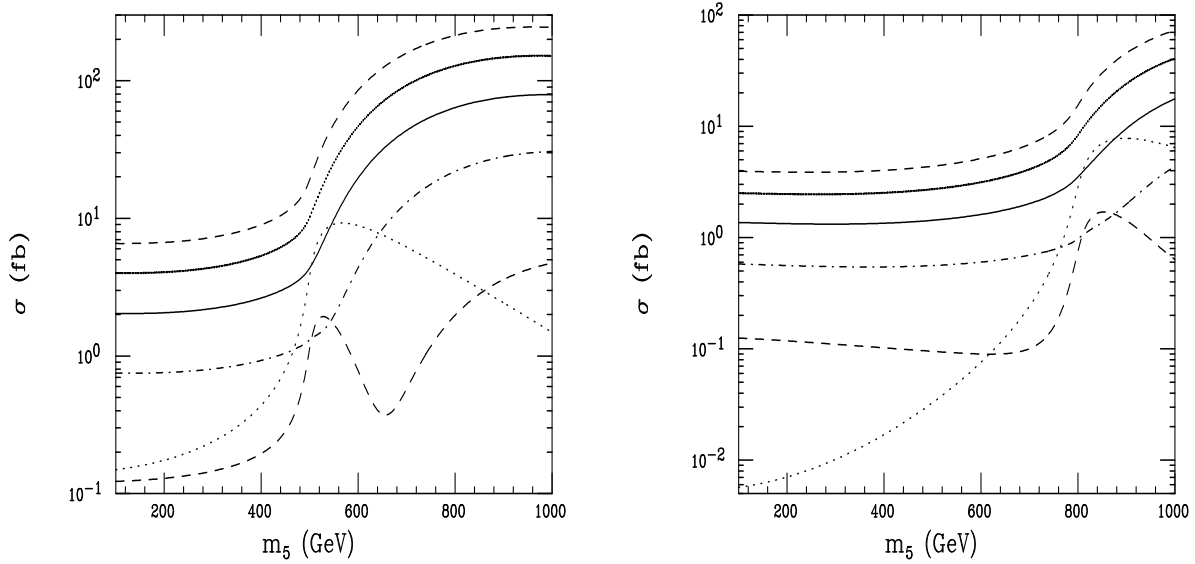

Fig. 6. $e^-e^- \to H_3^- H_3^- \nu\nu$ production cross sections as functions of $m_5$ with $c_H^2$ varying from 1 to 0 in steps of 0.2 from top to bottom on the left side of each figure. A 2 TeV NLC and $m_1 = m_{1'} = 250$ GeV is assumed in both cases. In (a)[(b)] $m_3 = 250[400]$ GeV is also assumed.